# Super-FEC Codes for 40/100 Gbps Networking

Zhongfeng Wang, *Senior Member*, IEEE

**Abstract** — *This paper first presents a simple approach to evaluate the performance bound at very low bit-error-rate (BER) range for binary pseudo-product codes and true-product codes. Then it introduces a super-product BCH code for optical transport networks (OTN). The code is shown to have very low error floor and can achieve near-Shannon limit performance with low decoding complexity.*

## I. Introduction

Modern communication systems are mostly speed hungry, *e.g.*, the Ethernet standard has evolved from 10 Mbps to 100Gbps in about two decades. When the target data rate hits the bottleneck of the transmission bandwidth, it is natural to consider complex modulation schemes, *e.g.*, DP-QPSK (dual-polarization quadrature phase-shift keying) proposed for 100Gbps optical transport networks (OTN). In these cases, it is essential to employ Forward Error Correction (FEC) code to lower the requirement of signal-to-noise ratio (SNR) at the receiver side.

In general, different FEC schemes can have significant differences in decoding performance, processing latency, and area/power consumption. Hence, an optimum FEC scheme is desired to well satisfy the requirements of the target application. For optical transport networking, it is generally required that the output bit-error-rate (BER) be below $10^{-15}$, the coding gain be close to channel capacity, and the codec be able to achieve very high data rate (e.g., > 40Gbps) with low cost.

It is known that both turbo codes and low-density parity check (LDPC) codes can achieve outstanding performance with moderate complexity [1]. But they both require soft inputs for iterative decoding. For ITU-T OTU2/OTU3/OTU4 optical transport networking, only hard-decision decoding is considered and the redundancy ratio is fixed at 6.7 % [2].

Conventional block codes such as RS codes and BCH codes are well-suited for hard-decision decoding. Additionally these codes have no error floor and the performance curve can be accurately computed for AWGN channel with BPSK modulation. In ITU-T G709, a standard RS(255, 239, t=8) code defined over GF($2^8$) was specified, where 255 denotes the total number of coded symbols, 239 is the total number of source symbols, t stands for error correction capacity.

However, such a simple code has a major drawback. The net coding gain (NCG) [2] at BER=$10^{-15}$ is only 6.2 dB. In principle, we can always improve the error correction capacity of a block code by increasing the code block size if the redundancy ratio is fixed. For instance, the RS (2720, 2550, t=85) defined over GF($2^{12}$), which was listed in ITU-T G975.1 specification [2], has a NCG of 8.0 dB. On the other hand, this long RS code requires many times more computational complexity than RS (255, 239, t=8) code in decoding.

An effective way to resolve the above conflict is to employ code concatenation [1], which results in pseudo-product codes such as one presented in [2] or true product codes. In ITU-T G.975.1 specification [3], only binary pseudo-product codes were considered, such as G.975.1.I4 code and G.975.1.I3 code. Product codes are normally decoded with iterative 2 phase decoding [1][4]. Good coding gains have been reported for these codes. However, those gains at very low BER range may be obtained through extrapolation of a simulated performance curve since normal software simulation cannot reach very low BER in a reasonable amount of time. When error floor appears, those extrapolations could deviate significantly from real performance curve.

In this work, a simple approach is first presented to evaluate performance bound of binary pseudo product codes, then a true product code with outstanding performance is introduced for OTN applications. Detailed performance analysis and comparison will be provided.

## II. Analysis for a Simple Example Code

First of all, let us take a simple pseudo-product code as an example. This code consists of 32 BCH(3908, 3824, t=7) codes as outer codes (or called row codes) and 64 BCH(2031, 1954, t=7) codes as inner codes (or called column codes). The outer codes are defined over GF($2^{12}$). The inner codes are defined over GF($2^{11}$). The parity bits for each row code are immediately appended after the corresponding source data. The parity bits of column codes are arranged at the end of corresponding column in the coded matrix. The detailed code structure is shown in Fig. 1. This code structure is similar to that of G975.1.I4, though the latter code has 16 Reed-Solomon codes instead of 32 BCH codes as row codes. The overall coded block has 129984 bits, where the source data have 122368 bits. The redundancy ratio is about 6.2%. This code is referred to as *Example Code-I* in later discussion.

It can be observed from Fig. 1 that each column code intersects with each row code for an overlapped segment of either 61 or 62 bits. If 8 or more bits errors occur in such an overlapped segment, the coded block will not be completely decodable since either row code or column code can only correct up to 7 bit errors. Such an error pattern is referred to as a *dead pattern*. Any dead pattern could set up a lower bound for output BER or upper bound for the NCG of the code at very low BER range. The probability of having such kind of dead patterns is computed with (1).

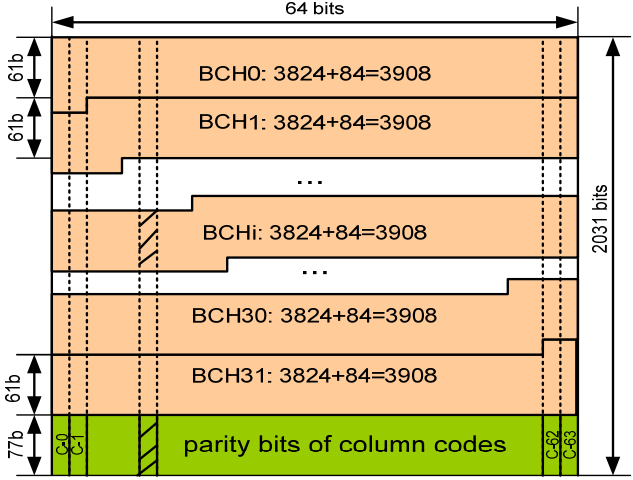

Fig. 1: Code structure of Example Code-I.

$$p_{11} = 1 - [1 - \sum_{i=8}^{62} \binom{62}{i} * e_{in}^{i} * (1-e_{in})^{62-i}]^{4*32}$$
$$* [1 - \sum_{i=8}^{61} \binom{61}{i} * e_{in}^{i} * (1-e_{in})^{61-i}]^{60*32} \quad (1)$$
$$> 1 - [1 - \sum_{i=8}^{61} \binom{61}{i} * e_{in}^{i} * (1-e_{in})^{61-i}]^{64*32},$$

where $e_{in}$ stands for the input BER to the decoder, the index 11 of p indicates that the dead pattern only involves the intersection between 1 row code and 1 column code. Apparently $p_{11}$ represents a lower bound for frame/block error rate (FER) for the FEC code. For simplicity, let us define $e_s$ as follows:

$$e_s = \sum_{i=8}^{61} \binom{61}{i} * e_{in}^{i} * (1-e_{in})^{61-i}. \quad (2)$$

So we have $p_{11} > 1 - (1-e_s)^{2048}$. It can be proved that, when $e_s * 2048 < 1$,

$$1 - 2048 * e_s < (1-e_s)^{2048} < 1 - 2048*e_s + \binom{2048}{2} * e_s^2.$$

Thus we have

$$p_{11} > 1 - (1-e_s)^{2048} > 1 - 1 + 2048*e_s - \binom{2048}{2}*e_s^2$$
$$= 2048*e_s - \binom{2048}{2}*e_s^2 \approx 2048*e_s \equiv \tilde{p}_{11}. \quad (3)$$

Since we are interested in very low BER range (e.g., $BER < 10^{-12}$), $\tilde{p}_{11}$ is very small and $e_s$ is even smaller. Thus it is reasonable to ignore $e_s^2$ term. If we define $\hat{e}_s = \sum_{i=8}^{I} \binom{61}{i} * e_{in}^{i} * (1-e_{in})^{61-i}$, let $I = 9$ to simplify the computation, and assume $e_{in} < 10^{-2}$, we have the following inequality:

$$p_{11} > 2048 * \hat{e}_s \equiv \hat{p}_{11}. \quad (4)$$

Thus we can compute a lower bound for the bit-error-rate caused by the above discussed error pattern using either $\tilde{p}_{11}$ or $\hat{p}_{11}$. Based on $\hat{p}_{11}$, we can compute the corresponding bit-error-rate:

$$q_{11} = 64*32*[\binom{61}{8}*e_{in}^{8}*(1-e_{in})^{53}*8$$
$$+ \binom{61}{9}*e_{in}^{9}*(1-e_{in})^{52}*9]/2031/64. \quad (5)$$

As can be observed from the above equations, we use *p* to denote a probability of an error event, and *q* to denote the corresponding bit-error-rate caused by this error event. For the estimation of BER bound, we are specifically interested in the range of $BER < 10^{-12}$. It can be derived that $\hat{e}_s < 10^{-11}$ and $e_s < 2*10^{-11}$ for this example code, which verifies $e_s*2048 \ll 1$. It can also be verified from (5) that $e_{in} < 0.01$ if the target BER<$10^{-12}$. Hence, it can be computed that the upper bound of NCG for this example code at BER=$10^{-15}$ is about 8.0 dB, which is significantly less than that projected from simulation curve (see Section IV). It should be pointed out that we can modify the *Example Code-I* so that its redundancy ratio reaches 6.7%, which is the same as the standard RS code used in OTN applications. In this case, all 32 row codes in the coded matrix are still BCH(3908, 3824, t=7) codes. But the first 52 column codes are BCH(2042, 1954, t=8) codes while the rest 12 column codes are BCH(2031, 1954, t=7) codes as before. The new code is denoted as *Example Code-IB*. The BER caused by Type-11 error patterns for this code can be computed as follows:

$$q_{11} \approx \{52*32*\binom{61}{8}*\binom{88}{1}*e_{in}^{9}*(1-e_{in})^{53}*8$$
$$+ 12*32*\binom{61}{8}*e_{in}^{8}*(1-e_{in})^{53}*8 \quad (6)$$
$$+ 64*32*\binom{61}{9}*e_{in}^{9}*(1-e_{in})^{52}*9\}/130560.$$

It is apparent that the modified code has slightly higher coding gain and lower BER bound. An important observation from above computations is that we can simply sum the BER contributed by each major dead pattern in estimating BER lower bound for a pseudo-product code at very low BER range. This is because that the probability of having two or more (identical or different) major error patterns appearing in a coded block at the same time is many orders of magnitudes lower than that of having a single error pattern. Hence, we will use the principle of simple summation for estimating BER bound in later discussion.

## III. PERFORMANCE ANALYSIS FOR GENERIC PSEUDO PRODUCT CODES

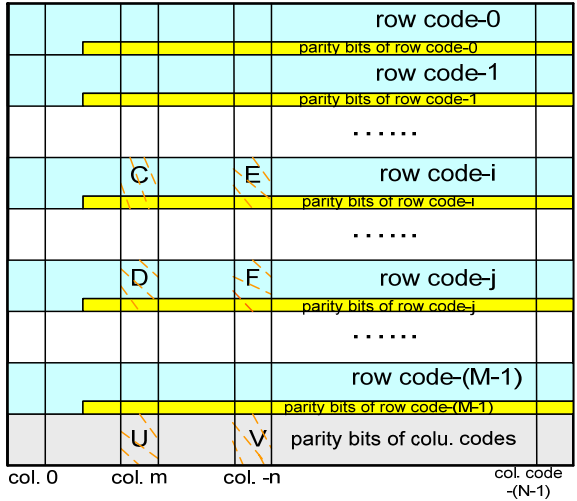

Fig. 2: Typical dead patterns for generic pseudo-product codes.

Now let us consider a general scenario. For convenience of ensuing analysis, we first make some simplifications for the code structure. We assume, with matrix row permutation, the parity bits of each row code are appended after its source data. We also assume each coded row code word covers an integer number of rows in the coded matrix. While these simplifications will greatly simplify the analysis of error bound, their impact to the final BER bound estimation should be negligible.

Fig. 2 shows the simplified code structure, where C, D, E, or F each represents an intersection between a row code and a column code. Both U and V stand for a parity bit portion of a column code. If an error pattern is only associated with one row code and one column code, then it is called Type-11 error pattern. If an error pattern involves two row codes and one column code, the error pattern is named as Type-21 pattern. Similarly we can define Type-12, Type-22, and Type-44 patterns, et al. For Type-11 error patterns, the corresponding BER can be computed using the following equation:

$$q_{11} = [\sum_i m_i \sum_{cond11} \binom{Ci}{j} * \binom{Ui}{k} * e_{in}^{j+k} * (1-e_{in})^{Ci+Ui-j-k} * j]/N_{in} \quad (7)$$

where $cond11 \equiv j+k > t_{ci}, j > t_{ri}$; $t_{ci}$ and $t_{ri}$ denote the error correction capabilities of corresponding row code and column code for the case i; $m_i$ denotes the total number of case i; Ci and Ui denote the total number of bits in the segments C and U respectively for the case i; $N_{in}$ denotes the total number of source data bits per code block.

For Type-21 error patterns, their resultant BER is computed as follows:

$$q_{21} = [\sum_i m_i \sum_{cond21} \binom{Ci}{j} * \binom{Di}{k} * \binom{Ui}{l} * e_{in}^{j+k+l} * (1-e_{in})^{Ci+Di+Ui-j-k-l} * (j+k)]/N_{in} \quad (8)$$

where $cond21 \equiv j+k+l > t_{ci}, j > t_{riC}, and\ k > t_{riD}$; $t_{ci}$, $m_i$, Ci, Ui, and $N_{in}$ are defined as before; $t_{riC}$ and $t_{riD}$ denote the error correction capabilities of corresponding row codes associated with segments C and D respectively for the case i; Di denotes the total number of bits in the segment D for the case i.

Computations for other types of error patterns can be derived similarly. For the estimation of BER bound, we are only interested in most likely (or major) dead (error) patterns. Thus, we will only list a few major error patterns in the following computations.

Next let us consider a pseudo-product code that is similar to G975.1.I3 code, but without interleaving. In this case, all 32 row codes are BCH(3860, 3824, t=3) codes. Column codes consist of 64 BCH(2040, 1930, t=10) codes. This code is denoted as *Example Code-II*. The BER caused by Type-11 error patterns is computed by using (9).

$$q_{11} \approx \{\binom{64}{1} * \binom{32}{1} * [\sum_{i=11}^{60} \binom{60}{i} * e_{in}^i * (1-e_{in})^{60-i} * i] + \binom{64}{1} * \binom{32}{1} * \sum_{i=4}^{10} \binom{60}{i} * \sum_{j=11-i}^{110} \binom{110}{j} * e_{in}^{i+j} * (1-e_{in})^{110+60-i-j} * i\}/122368. \quad (9)$$

For Type-21 error patterns, we only consider some major cases as following:

$$q_{21} \approx \{ \binom{64}{1} * \binom{32}{2} * 2 * [\binom{60}{5} * \binom{60}{6}$$
$$+ \binom{60}{4} * \binom{60}{7}] * e_{in}^{11} * (1-e_{in})^{109} * 11$$
$$+ \binom{64}{1} * \binom{32}{2} * \binom{60}{4} * \binom{60}{4} * \binom{110}{3}$$
$$* e_{in}^{11} * (1-e_{in})^{112} * 8 \} / 122368 \quad . \tag{10}$$

It is found that the above 2 types of error patterns lead to a lower BER bound of $10^{-15}$ when Eb/No ~ 6.9 dB. So this code has a NCG of less than 8.1 dB at the target BER, which is also much less that projected from simulation curves.

From the above analysis, we may conclude that a pseudo-product code generally has an error floor problem. A simple approach to improve the code performance is to linearly increase the block size. In particular, we can linearly increase the total number of row codes and the total number of column codes. Let us denote the double sized *Example Code-II* as *Example Code-IIb*. We have the following computation:

$$q_{11} \approx \{ \binom{128}{1} * \binom{64}{1} * [\sum_{i=11}^{30} \binom{30}{i} * e_{in}^{i} * (1-e_{in})^{30-i} * i]$$
$$+ \binom{128}{1} * \binom{64}{1} * \sum_{i=4}^{10} \binom{30}{i} * \sum_{j=11-i}^{110} \binom{110}{j} * e_{in}^{i+j}$$
$$* (1-e_{in})^{110+30-i-j} * i \} / 130560 / 2. \tag{11}$$

$$q_{21} \approx \{ \binom{128}{1} * \binom{64}{2} * 2 * [\binom{30}{5} * \binom{30}{6}$$
$$+ \binom{30}{4} * \binom{30}{7}] * e_{in}^{11} * (1-e_{in})^{49} * 11$$
$$+ \binom{128}{1} * \binom{64}{2} * \binom{30}{4} * \binom{30}{4} * \binom{110}{3}$$
$$* e_{in}^{11} * (1-e_{in})^{52} * 8 \} / 130560 / 2. \tag{12}$$

It will be shown later that the NCG bound of *Example Code-IIb* is 0.3dB larger than that of *Example Code-II* at target $BER=10^{-15}$. From simulations, we know the *Example Code-IIb* has an extra coding gain of less than 0.2dB over *Example Code-II* in the waterfall region. Further simulations indicate that the NCG bound generally increases much more than the NCG increased in waterfall region if we linearly increase the size of a pseudo-product code by increasing the total number of component codes. The extreme case of this kind of code size increase is when each row code covers one entire row and each column code covers one column in the coded matrix, which results in *a true product code*.

## IV. A TRUE PRODUCT CODE AND PERFORMANCE COMPARISON

Next we provide some analysis for a special true (binary) product code: the SP-BCH (called *super-product BCH*) code proposed by Broadcom in April 2009 to ITU-T for OTU4 100 Gbps applications [3]. The code consists of 960 BCH(987, 956, t=3) (1-bit extended) codes as row codes and 987 BCH(992, 960, t=3) (2-bit extended) codes as column codes. The detailed code structure is shown in Fig. 3. The most likely dead pattern is determined to be the 4x4 square error pattern (refer to Fig. 3), where 16 bits errors are located in the cross points between 4 arbitrary row codes and 4 arbitrary columns codes. Similar to computation in Equation (4), the bit error rate caused by this kind of dead patterns can be computed as follows:

$$q_{44} \approx \binom{992}{4} * \binom{987}{4} e_{in}^{16} * 16 / 970220 \tag{13}$$

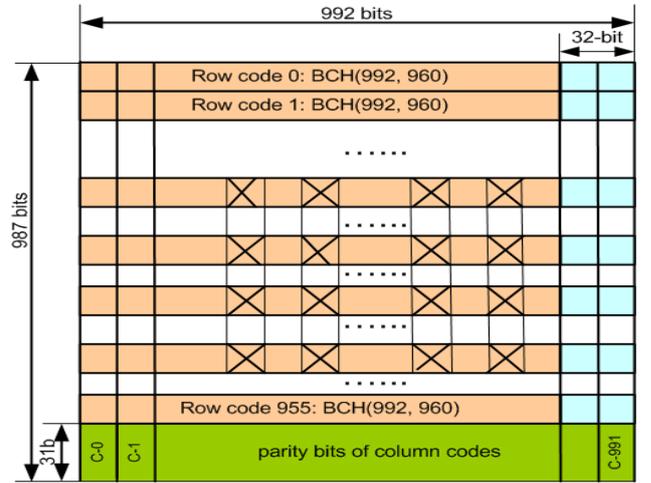

Fig. 3: Code structure of the SP-BCH code.

Fig. 4 shows performance comparison of several FEC codes: i) G709 standard RS(255, 239) code, ii) *Example Code-I*, iii) *Example Code-II,* iv) *Example Code-IIb,* and v) the SP-BCH code. All these codes have a redundancy ratio of 6 ~ 7 %. Only hard-decision decoding was considered in the simulation and performance bound computation. The *Example Code-I* were iteratively decoded for 4 iterations. The SP-BCH code was iteratively decoded for a maximum of 7 iterations assisted with a special technique called *dynamic reverting*.

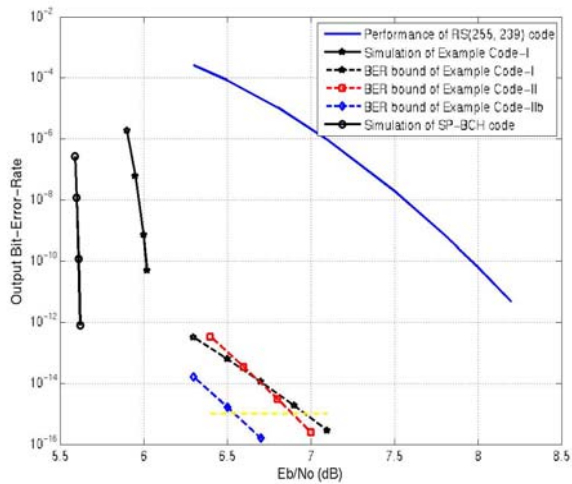

Fig. 4: Performance comparisons for four FEC codes.

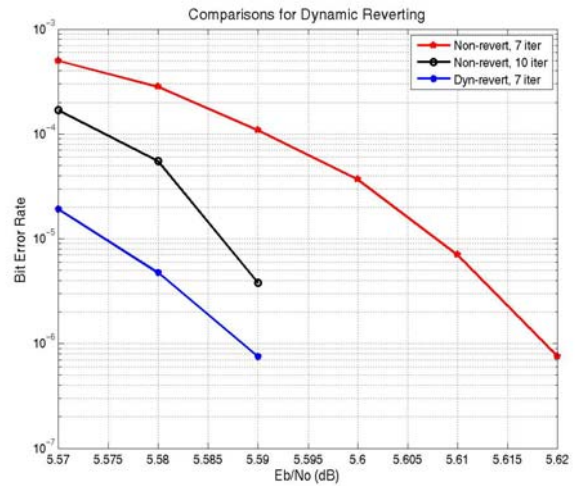

Fig. 5: Performance comparison for dynamic reverting

The *dynamic reverting* is an advanced decoding method. The key idea is to dynamically revert some of the decisions (*i.e.,* the corrected bits) from previous decoding. For instance, in the i-th row decoding phase, some of bits corresponding to j-th column code are corrected. In the following column decoding phase, if the j-th column code is determined as undecodable, all or part of the bits in the column corrected during the previous row decoding phase will be reverted. Similarly dynamic reverting can be employed during next row decoding phase, where some of the decisions made during the current column decoding phase may be reverted. Intuitively, when a false decoding happens, the decoded results will adversely affect the decoding of next decoding phase. So it is desired to change the decisions made by those false decoding. Hence, dynamic reverting is aimed to mitigate the effect of false decoding.

Fig. 5 shows performance comparison of decoding SP-BCH code with and without dynamic reverting. It can be observed that using dynamic reverting with 7 iterations can achieve the same or better performance compared to using conventional 2-phase iterative decoding with 10 iterations.

Moreover, it can be noted from Fig. 4 that the *Example Code-II* has lower error floor than *Example Code-I* while having similar coding gain in waterfall region (not shown in the figure), and the SP-BCH code has much higher coding gain in general over all other codes.

The SP-BCH code has been shown [3] to be a near-Shannon-limit code and is projected to achieve a BER of $10^{-15}$ when Eb/No is close to 5.6 dB (*i.e.,* NCG ~ 9.4 dB). On the other hand, at such a SNR, the BER contributed by the above error pattern is much less than $1.0 * 10^{-21}$, which means the SP-BCH code has an error floor below $10^{-21}$. This indicates we may encounter one undecodable block if we have 100 codec's to continuously run at 100Gbps for several years. It should be pointed out that some simple techniques (e.g., macro-level erasure decoding) can be used to eliminate the above mentioned dead pattern, which will bring the error floor many orders of magnitudes lower.

The major shortcoming for the SP-BCH code is its relatively long decoding latency due to large block size. However, it should be acceptable for 40Gbps and higher data rate applications. Moreover, it has some other superior advantages. For instance, its computational complexity (per bit) is much less than that of *Example Code-*I or G975.1.I4 code since, for t=3 BCH code, it is feasible to use direct equation solver to find out error locations. In addition, the required maximum number of iterations for this SP-BCH code is significantly less compared to *Example Code-I* or any other code with similar NCG for a practical application since the former code has much higher coding gain. Also the burst error correcting capacity of the SP-BCH code is nearly 3000 bits without any advanced decoding, which is 6 times better than that of *Example Code-I*. If the above mentioned erasure decoding is employed, the burst error correcting capacity of SP-BCH code can reach nearly 7000 bits, which is many times larger than any existing code for network transport applications. In particular, the SP-BCH code is amenable to VLSI implementation. It has been demonstrated that an entire codec of SP-BCH code can be implemented on a single Altera FGPA device to deliver over 40 Gbps data rate. However, it will be a great challenge to design a single codec for *Example Code-I, Example Code-II*, or any other similar code to achieve similar throughput with one FPGA device. In brief, although the SP-BCH code has drawback in decoding latency, its many advantages renders it an optimal option for FEC code used for 40Gbps and beyond high-speed network applications.

## V. CONCLUSION

In this work, we have proposed an efficient way for performance evaluation at very low BER range for binary pseudo-product and true product codes. The key idea is to identify most likely dead patterns for the target product code. We have also introduced the SP-BCH code, a true-product code and an advanced decoding method named dynamic reverting, which demonstrated outstanding performance while owing very low decoding complexity.

### ACKNOWLEDGMENT

The author is very much grateful to Dr. G. Ungerboeck for his seminar work on performance analysis of product codes at Broadcom in 2009.

### REFERENCES

[1] S. Lin and D. Costello, "Error Control Coding", $2^{nd}$ edition, Prentice Hall, 2004.
[2] K. Lee, et al, "100GB/S two-iteration concatenated BCH decoder architecture for optical communications," IEEE Workshop on Signal Processing Systems, pp. 404-409, Oct. 2010.
[3] ITU-T G.975.1, "Forward error correction for high bit-rate DWDM submarine systems," available from http://www.catr.cn/radar/itut/201007/P020100707621029951507.pdf.
[4] S. Changuel, "Iterative decoding of product codes over binary erasure channel," Electronics Letters, vol. 46, no. 7, Apr. 2010.
[5] Z. Wang, etc, "Communication device employing binary product coding with selective additional cyclic redundancy check (crc) therein." US patent application, Apr. 2010.

### BIOGRAPHIES

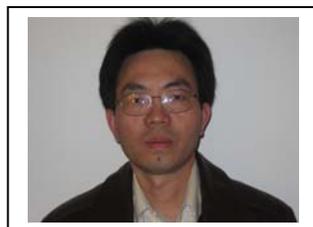

**Z. Wang** (M'00-SM'05) received both B.E. and M.S. degrees from Tsinghua University, Beijing, China. He obtained his PhD degree from the University of Minnesota in 2000. In the past, he has worked in both academia and industry. From 2003 to 2007, he was an assistant professor in the School of EECS at Oregon State University. He is currently an Associate Technical Director at Broadcom Corporation, Irvine, California. He has edited one book "VLSI" (InTech Publisher), authored/coauthored over 100 technical papers, and filed tens of patent applications and disclosures. He was the recipient of IEEE Circuits and Systems Society VLSI Transactions Best Paper Award in 2007. He is also a coauthor of three papers among top ten most downloaded manuscripts in IEEE Transactions on VLSI Systems from 2007 to 2009. He has been serving as Associate Editor for IEEE Transactions on Circuits and Systems-I (2003-04), IEEE Transactions on Circuits and Systems-II (2008-13), and IEEE Transactions on VLSI Systems (2009-12). His current research interest includes VLSI design for high-speed networking systems.